%% file: Journal_paper.tex
\documentclass[letter, draftcls, onecolumn, 12pt]{IEEEtran}
\usepackage[noadjust]{cite}
\usepackage[nolists, nomarkers]{endfloat}
\usepackage{latexsym}
\usepackage[latin1]{inputenc}
\usepackage[T1]{fontenc}
\usepackage[dvips]{graphicx}
\usepackage{verbatim}
\usepackage{amsmath}
\usepackage{amsfonts}
\usepackage{amssymb}
\usepackage{exscale}
\usepackage{latexsym}
\usepackage{bm}

\title{On the Behavior of RObust Header Compression U-mode in Channels with Memory}

\author{
Romain Hermenier, Francesco Rossetto, Matteo Berioli\\
\thanks{The material in this paper was presented in part at ISWCS 2011, Aachen (Germany). Romain Hermenier and Matteo Berioli are with the Institute of Communications and Navigation, German Aerospace Center (DLR), Oberpfaffenhofen, 82234 Weßling, Germany. (e-mail: {Romain.Hermenier; Matteo.Berioli}@dlr.de). Francesco Rossetto was with DLR, he is now with Rohde \& Schwarz GmbH \& Co., Mühldorfstraße 15, 81671 Munich, Germany (e-mail: Francesco.Rossetto@ieee.org).}
}

\begin{document}

\setcounter{page}{0}

\maketitle
\baselineskip 22.5pt

\begin{abstract}
The existing studies of RObust Header Compression (ROHC) have
provided some understanding for memoryless channel, but the behavior
of ROHC for correlated wireless channels is not well investigated in
spite of its practical importance. In this paper, the dependence of
ROHC against its design parameters for the Gilbert Elliot channel is
studied by means of three analytical models. A first more elaborated
approach accurately predicts the behavior of the protocol for the
single RTP flow profile, while a simpler, analytically tractable
model yields clear and insightful mathematical relationships that
explain the qualitative trends of ROHC. The results are validated
against a real world implementation of this protocol. Moreover, a
third model studies also the less conventional yet practically
relevant setting of multiple RTP flows.
\end{abstract}

\newpage

\input{Introduction.tex}
\input{System_Description.tex}
\input{Model.tex}

\input{Numerical_Results_Validation.tex}
\input{Conclusions_Acknowledgments.tex}
\bibliographystyle{IEEEbib}
\bibliography{IEEEabrv,./biblio.bib}

\end{document}

%% file: Introduction.tex
\section{Introduction}\label{sec:Introduction}

The concept of header compression has been applied very successfully
in the wired world and has lead to very effective compression of
long IP headers. Since the headers of two consecutive IP packets are
highly correlated, the essential idea is to transmit the
compressed version of the difference between these headers. This
compression can be very effective, but is also fragile to packet
losses. While such losses do not frequently happen over wires, they
are far more common for the wireless medium. Hence, the traditional
header compression mechanism is inadequate to withstand these error
ratios and ROHC has been introduced~\cite{Bormann01, Taylor05,
Pelletier06, effnet04}.

The ROHC protocol has been developed and standardized by the IETF in
2001 and aims at reducing the header sizes of IP packets to be sent
through a cellular link. It offers a strong resilience against
channel losses and yet a high compression efficiency. The scheme is
so effective that it has found its way in important wireless
standards like HSPA and LTE~\cite{effnet04, West02, Fitzek05}, and is being currently proposed for the next generations of DVB RCS and DVB SH. The
majority of the past studies are simulation-based, and investigate
the performance of ROHC in different
environments~\cite{Taylor05, Fitzek05, Wang04a, Woo08, Fortuna05,
Piri08, Jin04}. On the other hand, only a limited amount of research
has tried to analytically describe ROHC~\cite{Kalyanasundaram07,
Cho05, Couvreur06, Wang04}. These models can be quite accurate but
they do not provide simple expressions and hence rarely do they
offer deep insight in the behavior of the protocol as a function of
important design parameters or of the channel characteristics.
Moreover, all models (except for~\cite{Cho05}) are derived for
memoryless channels and for traffic with just a single RTP/UDP/IP flow,
while in fact the wireless medium is often correlated in time and
ROHC leads to interesting capacity improvements especially if
several flows are multiplexed together~\cite{Piri08}.

Our investigation has focused on the performance of ROHC in
correlated channels, in particular the Gilbert-Elliott
one~\cite{Badia10}.
The core contribution of our work lies in the development
of three analytical models for ROHC. The first one
accurately mimics the effective behavior and performance of
single-RTP-flow ROHC with the Gilbert-Elliott channel. The second
model is a simplified version of the first one that still fairly predicts the trends of
ROHC but can be solved in closed form and yields simple and
insightful formulae. These results enable to draw useful
relationships between the system performance and the design
parameters, like the timeout for the transmission of the
uncompressed headers or the interpretation window. The third model
explores the effect of multiplexing several RTP/UDP/IP flows together on the
system performance. Our contribution serves two main purposes: on
the one hand, to better characterize the performance of ROHC in
correlated channels, on the other hand to provide simple and useful
relationships for the design and tuning of a ROHC system.

The rest of the paper is organised as follows. Section~\ref{sec:System_Model} introduces the system model and recalls the most relevant properties of ROHC for the present discussion. Section~\ref{sec:Model} describes the three proposed analytical models, which represent our main contribution. The trends and predictions are numerically studied in Section~\ref{sec:Numerical_Results}, while Section~\ref{sec:Conclusions} draws the conclusions.


%% file: System_Description.tex
\section{System Model}\label{sec:System_Model}

A slotted system with a single transmitter is considered. Without
loss of generality, a setting with only one receiver is studied. The
transmitter (or source, in the rest of this paper) can either generate one single packet stream for which a ROHC profile between 1 and 4 applies or multiplexes
$M$ different RTP/UDP/IPv4 flows. In the former case, the applicable protocols are RTP, UDP, ESP and IP. It is assumed that the source is saturated
and has an available packet for transmission in every slot.

The channel is regarded as a Gilbert Elliott packet deletion
channel~\cite{Badia10}. This channel is modeled by means of a
two-state Markov chain: the good state G (correct reception of the
packet) and the bad state B (the packet is lost and the upper layers
are not aware that a packet was sent). Let us define as
$\mathcal{P}$ the one step transition matrix of the Markov chain and
as $P_{\text{X,Y}}$ the transition probability from state $X$ to
$Y$, $X,Y\in\{\text{G},\text{B}\}$. The transition matrix is
uniquely determined by $P_{\text{G,B}}$ and $P_{\text{B,G}}$, which
are inversely proportional to the average time spent in the good and
bad state, $L_\text{G}$ and $L_\text{B}$, respectively. The
Gilbert-Elliott channel is also equivalently defined by the average
duration of a sequence of consecutive bad states
$L_\text{B}=1/P_{\text{B,G}}$ and the average deletion probability
$\epsilon=P_{\text{G,B}}/(P_{\text{G,B}}+P_{\text{B,G}})$~\cite{Badia10}.
We remark the assumption of packet deletion, rather than erasure~\cite{Zigangirov69}. It
turns out that ROHC does not assume that the lower layers would
provide a feedback to the IP and above layers in case the packet was
not successfully decoded. Hence, the packet is either correctly
received or the ROHC receiver is simply unaware of the loss and therefore the terms "packet loss" and "packet deletion" will be used interchangeably in the rest of this paper. Finally
we also remark that this channel model holds for consecutively transmitted
packets, by other words it is satisfactory for the single RTP
flow setting ($M=1$). The necessary modifications for $M>1$ will be
explored in Section~\ref{subsec:Multiple_Flows}.

In the rest of this section, a quick introduction on ROHC and on the
investigated elements shall be provided. Three different modes of
operation can be used in ROHC: Unidirectional (U), Bidirectional
Optimistic (O) and Bidirectional Reliable (R). The major difference
between these three modes is how the state transitions are handled
and the lack of a feedback channel for the U-mode. Moreover every
protocol (IP, UDP, RTP) is linked to a specific configuration of
ROHC called profile~\cite{Bormann01}. The focus within this paper is
on the U-mode  and on the RTP/UDP/IP, UDP/IP, ESP/IP and IP profiles (i.e., profiles 1 to 4, respectively). We
refer to~\cite{Bormann01} for a detailed description of the O-mode
and R-mode.

\begin{figure}[!ht]
\centering
\includegraphics[width=0.8\columnwidth]{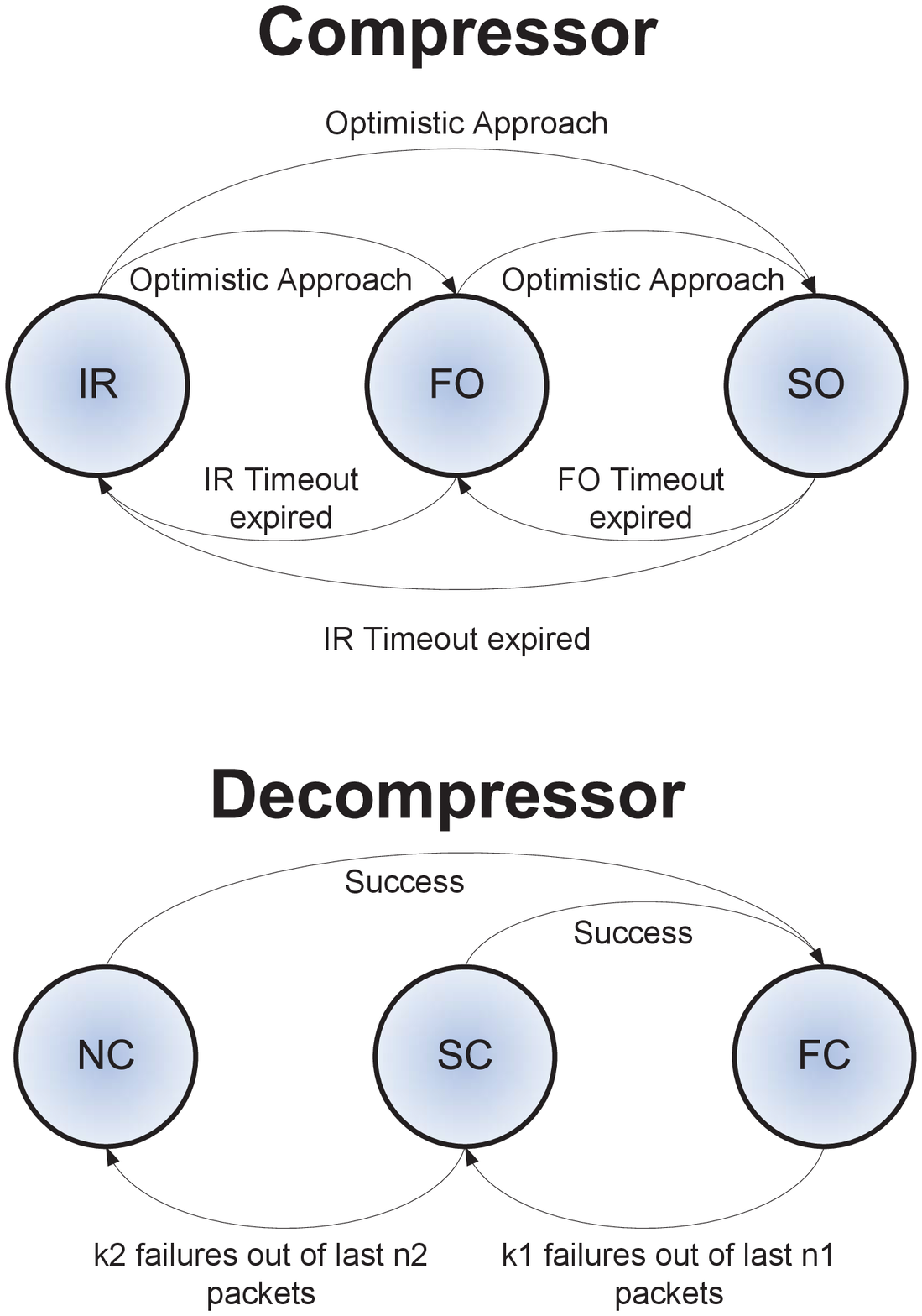}
\caption{ROHC state machines of the compressor and decompressor in U-mode.} \label{fig:uni_state_machines}
\vspace{-5pt}
\end{figure}

ROHC has two main key properties, namely its efficiency and its
robustness. ROHC achieves a very high compression efficiency thanks
to properly tuned state machines at the compressor and decompressor
sides, which are based on the fact that the fields of a packet
header are classified into two categories: static (such as UDP Port numbers) and dynamic (such as the Hop Limit in IPv6 or the Identification in IPv4). The compressor sends first
uncompressed packets called Initialization and Refresh (IR) packets,
so that both the compressor and the decompressor can initialize
their context by storing information concerning the header. Once
the compressor is confident enough that the decompressor
successfully received an IR packet (by use of ACK for O- and R-mode
or by sending $L$ IR packets for the U-mode), it switches forward to
an intermediate compression state (First Order (FO), where
only the dynamic fields of the header are sent uncompressed) or
directly to a full compression state (Second Order (SO), where the
header is entirely compressed) as displayed in
Fig.~\ref{fig:uni_state_machines}. In U-mode two different timeouts are used
to periodically switch downward to less efficient compression states
(IR and FO Timeouts). These two timeouts ensure the context
synchronization between the compressor and decompressor since no
feedback channel is considered in U-mode. Finally transitions from IR to more compressive states work based on the optimistic approach as shown in Fig.~\ref{fig:uni_state_machines} and explained in~\cite{Bormann01}.

The state machine at the decompressor side is also composed of three
states. After the initialization of the context due to the good
reception of an IR packet, the decompressor moves from the No
Context (NC) state where only IR packets can be decoded to the Full
Context (FC) state where all kinds of ROHC packets (IR, FO, SO) can
be decompressed (Fig.~\ref{fig:uni_state_machines}). The decompressor
switches from FC to SC only if $k_{1}$ packets out of the last
$n_{1}$ received packets have been unsuccessfully decoded (CRC
failed).\phantom{\scriptsize{l}}\footnote{Unless otherwise stated, by CRC it is meant the one introduced by ROHC to check the correctness of the reconstructed header.} In this intermediate state (Static Context (SC)) the
decompressor can only decode IR or FO packets. Therefore if it
receives one of them and the decompression is successful, it moves
back to the FC state. However, if over the last $n_{2}$ received
packets, $k_{2}$ had a CRC failure, the decompressor moves downward
to the NC state, where it will wait for an IR packet (all other
received packets in this state are dropped). We refer
to~\cite{Bormann01, Hermenier01, effnet04, Couvreur06} for further
information.


The second key property of ROHC is its ability to resist to larger
packet error ratios than classic header compression schemes. This
very high robustness is achieved by the combined use of an encoding
scheme and of a second algorithm which is employed when too many
consecutive packets are lost~\cite{Hermenier01}. The starting point is the fact that a SO packet of profiles 1 to 4 includes a compressed version of a suitable sequence number (e.g., the RTP SN in profile 1 or the UDP SN generated by ROHC in profile 2). The encoding scheme
called Window-based Least Significant Bits (W-LSB) is defined by an
interpretation interval $[-p, 2^{k}-1-p]$ of size $2^{k}$, where $k$
represents the $k$ least significant bits of the encoded field value
and $p$ the offset with respect to the previously received field
value~\cite{Bormann01, Cho05}. If the SN of the received packet belongs to the interpretation interval, the header can be successfully decompressed and the context updated. Therefore, up to $2^k-1-p-1$ packets can be lost in a row without losing synchronization between compressor and decompressor, since field values undergoing small negative changes are not considered here. The second algorithm is called LSB wraparound and enhances the robustness of ROHC by
shifting the interpretation interval of $2^{k}$ when more than $2^k-1-p-1$
consecutive packets are lost~\cite{Bormann01}. Thus, the maximal
number of packets that can be deleted in a row while still retaining
context synchronization can be defined as~\cite{Hermenier01}:

\begin{equation}\label{eq:W_parameter}
   \vspace{-5pt}
    W = (2^{k}-1-p)-1+2^{k}=2^{k+1}-2-p
\end{equation}

If more than $W$ packets are lost in a row, the decompressor is not
able to decode the next arriving SO packet and is said to be
out-of-synchronization. When the receiver is out-of-synchronization
and in FC or SC state, the reception of an IR or FO packet enables
to retrieve the synchronization, whereas in the NC state the
decompressor only updates its context by means of an IR packet.

\begin{figure}[!ht]
\centering
\includegraphics[width=0.8\columnwidth]{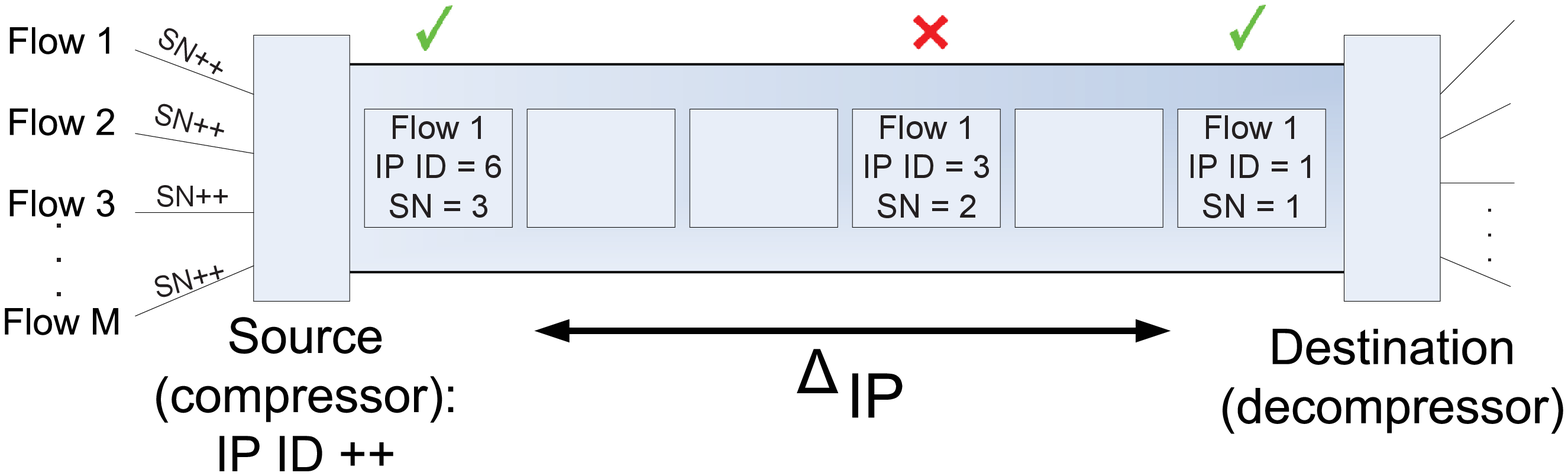}
\caption{$M$ different RTP flows multiplexed using the same compressor and decompressor} \label{fig:mult_flows}
\end{figure}

If several RTP/UDP/IPv4 flows are multiplexed through the same compressor, the
IP identifier is increased by one for each outgoing packet from the
source~\cite{Bormann01}. Focusing on one of these RTP flows, the IP
identifier will not increase linearly since packets from other RTP
flows may be inserted between two packets from the tagged flow.
Fig.~\ref{fig:mult_flows} shows an example where the focus is on the
first flow. Hence, the IP identifier can not be retrieved from the
SN and must be sent in the compressed header~\cite{Bormann01}.
According to~\cite{Bormann01}, the same W-LSB encoding scheme is used to
encode the IP identifier field, although with different values of $k$ and $p$
with respect to the ones used to encode the SN. Thus, the interpretation interval is equal to $[-p, 2^{k_o}-1-p_o]$. In addition, since no wraparound
algorithm is used~\cite{Bormann01}, $W_o$ can be defined as the upper limit of this interpretation interval:

\begin{equation}\label{eq:Wo_parameter}
   \vspace{-5pt}
    W_o = (2^{k_o}-1-p_o)-1
\end{equation}

This means that if the IP
identifier increases by more than $W_o$ between two consecutively
correctly received packets of the same RTP flow, the decompressor
will not be able to decode the next packets since too many IP
packets from other RTP flows have been inserted.

Thus for a single flow communication ($M=1$) $W$ is the only key
parameter for the definition of the ROHC robustness, whereas in the
case of multiple RTP sources the robustness of the header compression
scheme is limited by:
\begin{itemize}
\item $W$, which gives the maximum number of packets from
the same RTP flow that can be lost in a row and still keeping the
context synchronization;
\item $W_o$, which specifies the maximum number of IP packets
that can be inserted between two consecutive packets from
the tagged RTP flow without losing the synchronization between
the compressor and the decompressor.
\end{itemize}

%% file: Model.tex
\section{Model Derivation}\label{sec:Model}


Out of the description of the ROHC protocol provided in Section~\ref{sec:System_Model}, an important observation can be made, which is at the basis of the first two models.

It happens in many circumstances that all fields of the header to be compressed can be inferred from just one field. For instance, this can be the case in Profile 1 for RTP/UDP/IP headers, when all dynamic fields have a constant known offset with respect to the RTP Sequence Number, whereas the static ones are already stored in the context~\cite{Bormann01}.\phantom{\scriptsize{l}}\footnote{Some fields like the RTP Marker bit change infrequently and are transmitted when necessary. However this fact does not change the gist of the described models.} In such cases, the Window LSB encoding method guarantees that the key field can be correctly retrieved as long as no more than $W$ headers are deleted in a row. Our model works for any ROHC profile for which the previous property holds. For example, profiles 1 to 4, where the sender side needs to compress only one header field (excluding the CRC) and the decompressor can retrieve all other fields from the compressed one. In these cases, if a single flow is present, the only field to encode is the RTP, UDP, ESP or IP Sequence Number (SN), respectively.\phantom{\scriptsize{l}}\footnote{We remark that the UDP, ESP and IP sequence numbers are introduced by the compressor and are not standard fields of the respective protocols.} The SN is incremented by one for each sent packet and is to be used by the receiver to detect packet losses and to reorder the packets.

In some other cases, more than one dynamic field can often change its value and hence two or more fields need to be compressed. Model 3 will explore one such examples for the specific case of RTP/UDP/IPv4 headers, when multiple RTP flows are multiplexed in the same IP flow. In this setting, the IP ID field ceases to have a constant known offset from the RTP SN and the compressor can resort, for instance, to offset encoding or scaled encoding~\cite{Bormann01}, Sections 4.5.3 and 4.5.5, respectively.

\subsection{Model 1: Single Flow, full representation} 

\begin{figure}[!ht]
\centering
\includegraphics[width=\columnwidth]{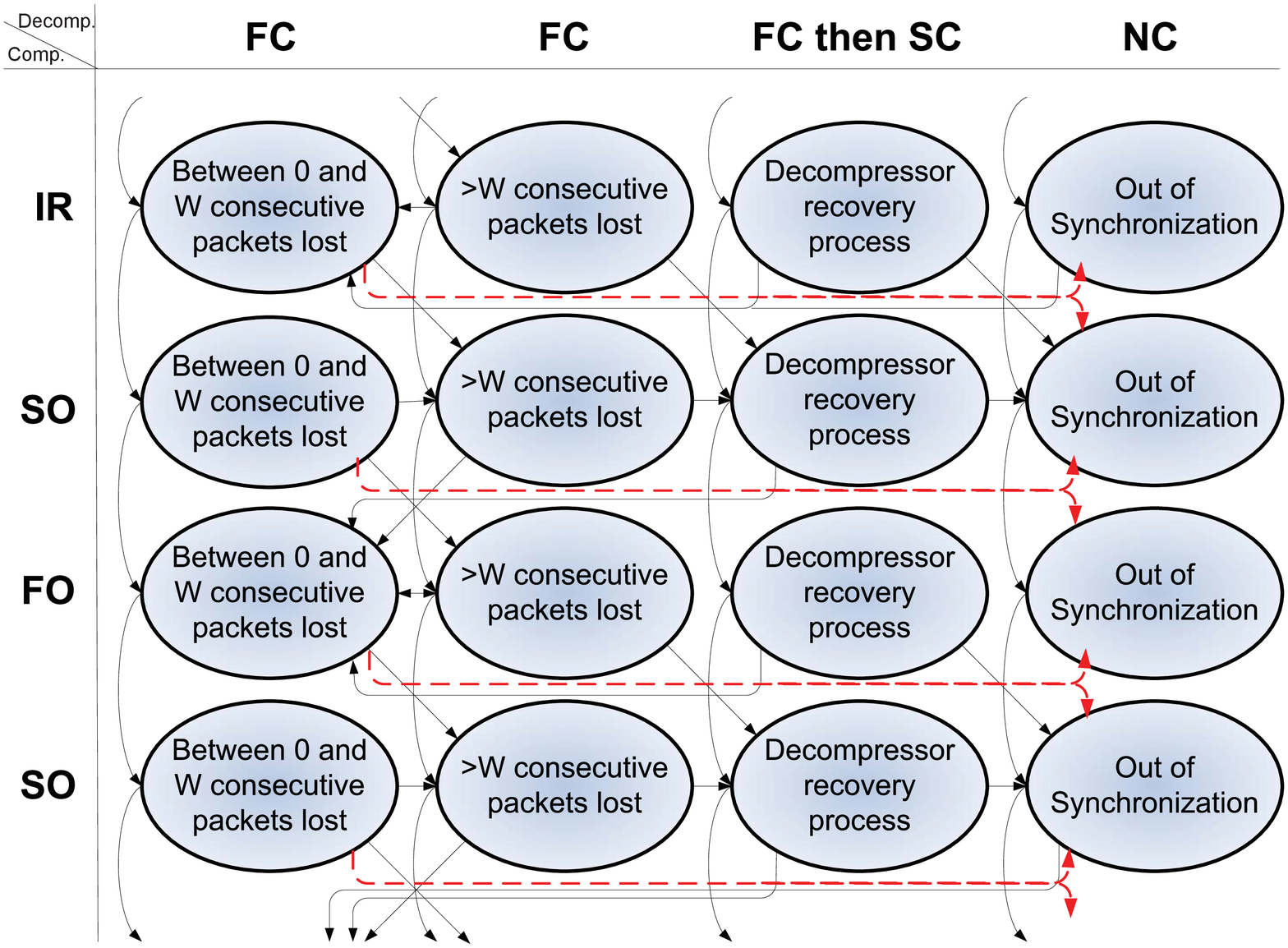}
\caption{Simplified Markov chain model for the ROHC modeling in U-mode - model 1} \label{fig:model1}
\end{figure}

On the basis of these observations, a realistic Markov chain fully
compliant with the ROHC standard has been derived (model 1). The
states of this Markov chain can be arranged in a two dimensional
array which tracks on the one dimension the number of packets lost
in a row and on the other dimension the kind of ROHC packets
transmitted by the compressor. Fig.~\ref{fig:model1} depicts a
simplified scheme of this Markov chain where the states for which
the compressor and the decompressor share the same state are
coalesced in ellipses. For a better understanding the status of the
state machines of the compressor and the decompressor have been
added.

Each of the ellipses displayed in Fig.~\ref{fig:model1} may contain
more than one state (up to a few hundred for some ellipses). This detailed model turns out to be accurate, but
is not amenable to
mathematical analysis due to its large size since it can reach
thousands of states depending on the configuration of the ROHC
parameters.
Each row of the array in Fig.~\ref{fig:model1} describes the kind of
ROHC packets that have been sent by the compressor. As explained in
Section~\ref{sec:System_Model}, when the decompressor is in FC or SC
state and more than $W$ packets have been lost consecutively, the
synchronization is retrieved upon a correct acquisition of an IR or
FO packet whereas only the correct reception of an IR packet enables
the decompressor to recover the synchronization when it is in NC
state. Thus the type of sent packets is a parameter to be compulsory
tracked.

The second parameter to be followed is the number of consecutively lost packets
(tracked by the columns of the matrix). Here the major
point to differentiate is whether more than $W$ packets have been
deleted consecutively or not (represented by the two first ellipses of
each line in Fig.~\ref{fig:model1}). In order to correctly model the
ROHC standard, an additional ellipse has been represented before the
out-of-synchronization state so as to map the behavior of the
decompressor upon the correct reception of a ROHC packet when more than
$W$ packets have been lost consecutively (third ellipse of each
line). According to~\cite{Bormann01} the synchronization can be
retrieved if a FO or IR packet is correctly received although more
than $W$ packets have been deleted in a row. This is due to the "$k$
out of $n$" rule used by the decompressor to recover a
packet~\cite{Bormann01, Hermenier01}. This rule explains the two
possible states of the decompressor for this ellipse as well ("FC
then SC"). Finally the last ellipse of each line in
Fig.~\ref{fig:model1} represents the out-of-synchronization state
when the decompressor can not decode any packet and is waiting for
the correct reception of an IR packet to retrieve the synchronization.
Since a Gilbert-Elliott deletion channel is considered, this ellipse
contains in reality two out-of-synchronization states: one for the
good state and one for the bad state.

A full description of the model cannot be provided due to reasons of space. However, the ellipse from the second row, first column of Fig.~\ref{fig:model1} will be further explained, as the inner structure of the other regions of the Markov chain is similar. This ellipse is composed by $W+1$ columns (each tracking a different value of $w$) and as many rows as the time between the expiration of the IR timeout and the transmission of the next FO header. Let us assume to be in a generic state of this ellipse. Every time a new header is sent, the chain transitions to the next row below. If the column is the first one, the channel must have been in the non-deletion state ($G$), otherwise it is in the deletion ($B$) state. The decompressor correctly receives the next header if the channel will not be in the bad state, which happens with probability $P_{\text{G,G}}$ and $P_{\text{B,G}}$, respectively, and will transition into the first column. Otherwise, the column index will increase by one. If $w$ was already $W$, the chain transition into the next ellipse to the right.

Moreover in order to be fully compliant with the ROHC protocol, the
errors due to a wrong CRC check in the robustness mechanism have
been introduced (represented by the dashed red arrows in
Fig.~\ref{fig:model1}). To better understand this issue let us write
$W$ as follows:

\begin{equation}\label{eq:W_parameter_details}
   \vspace{-5pt}
    W = W_1 + W_2
\end{equation}

where

\begin{eqnarray}
    W_1 &=& (2^{k}-1-p)-1 \label{eq:W1_parameter} \\
    W_2 &=& 2^{k}         \label{eq:W2_parameter}
\end{eqnarray}

As explained in Section~\ref{sec:System_Model}, if more than $W_1$
but less than $W_2$ packets are lost in a row, ROHC applies the
wraparound algorithm to enhance its robustness. However, before
applying this algorithm, ROHC tries a first time to decode the
received packet and performs a CRC check. Since more than $W_1$
packets have been lost consecutively, ROHC can not retrieve the
received packet without the wraparound algorithm and the CRC check
is therefore wrong. As the decompressed packet did not pass the CRC
check, ROHC applies the wraparound algorithm and tries to decode the
packet again, but this time the interpretation interval is shifted
and hence a new header is reconstructed. If the new CRC check is
successful the packet will be forwarded to the higher layers. As explained in the Appendix, the CRC at the first attempt may nonetheless yield a false
negative with a probability of 1/32 (i.e., the packet is wrong but
the CRC check did not realize it). In this case, the wrong packet is
regarded as valid and is forwarded to the upper layers which will
not be able to interpret it. This wrong interpretation of the CRC
explains why the chain can switch directly from the first ellipse of
each line to the out-of-synchronization state.

The probability of losing synchronization between the compressor and
decompressor can be numerically computed as the sum of the steady state probabilities of being in one "OoS" state (see the last column of Fig. 3). This
out-of-synchronization (OoS) probability only depends on the ROHC
design parameters (L, IRT) as well as on the characteristics of the
wireless channel. The accuracy of this model makes it suitable to
provide a quick yet faithful evaluation of the protocol performance.
However, this model offers neither a simple relationships between
the above mentioned metrics nor deep insight into the protocol
behavior.

\subsection{Model 2: Single Flow, simplified representation}\label{subsec:single_simplified} 

The model that has been described in the previous subsection
faithfully represents the behavior of the ROHC protocol and
Section~\ref{sec:Numerical_Results} will prove the very good match
between predictions and real world measurements. The previous model
is however quite elaborated and does not enable to derive simple
relationships of the system performance. The goal of the second part
is to devise a model with far fewer states, that yields
clear closed form expressions for the OoS probability and hence
more insight on the behavior of ROHC, yet at no major loss of
modelling accuracy.

In order to characterize the system, three elements must be
modelled: the compressor, the channel, and the decompressor. The
model described within this subsection (model 2) is based on a set
of assumptions, which aim to reduce the complexity of model 1 while
still correctly predicting the qualitative trends of the protocol
performance against the design parameters. These assumptions are
listed hereafter:

\begin{itemize}
\item \emph{Channel model}: A Gilbert-Elliott packet deletion channel is considered, as stated in Section~\ref{sec:System_Model}.

\item \emph{Compressor}: The FO packets are not taken into account for this model because of their limited actual impact. Moreover, $L=1$ for the sake of simplicity, while $L$ is arbitrary in model~1. The compressor state machine comprises only two states: IR and SO states. Moreover, it is assumed that the compressor decides the type of the packet between IR and SO independently in every slot. An IR frame is generated with probability $P_{\text{IR}}$, and therefore the number of SO frames between two IR packets (i.e., the IR timeout) follows a memoryless, geometric random variable with average value:

\begin{equation}\label{eq:proba_ir}
   \vspace{-5pt}
\text{IRT}=\frac{1}{P_{\text{IR}}}
\end{equation}

\indent By means of these assumptions, the IRT is no longer deterministic
and becomes geometric. Thus the knowledge of the compression
level of the previous packets is not required anymore. We remark that some models of other systems which employ backoffs to resolve collisions also assumed a geometric backoff to yield more tractable formulae, even though such quantity can have other distributions in practical realisations of these systems (see for instance the analysis of IEEE 802.11
\cite{Cali00} or of ALOHA~\cite{Lam74}). The qualitative trends
of the system are still correctly predicted, while the numerical
performance is often about the same up to a multiplicative
constant. Section~\ref{sec:Numerical_Results} provides a
validation of this claim against a real world measurement.

\item \emph{Decompressor}: Since no FO packets are considered, the SC state of the
decompressor is omitted as well. Therefore the decompressor
state machine is composed of two states (the NC and the FC
states). The key element to track is the number of consecutively
lost packets. The following modelling assumptions for the
decompressor have been included:

\begin{itemize}
\item If no packets are lost, the decompressor remains in FC state and works properly.
\item If no more than $W$ packets are lost in a row due to channel impairments,
the decompressor is still able to decode the next SO packets
thanks to the ROHC algorithm.
\item However if the decompressor realises that more than $W$ packets are lost, there is a major
context damage and the decompressor cannot interpret the
next arriving packet. Thus it switches back directly to the
NC state and waits for an IR packet. Neither the $k_{1}$ out
of $n_{1}$ nor the $k_{2}$ out of $n_{2}$ rules are
considered. Until the correct
reception of an IR packet, the decompressor is
out-of-synchronization.
\end{itemize}

\end{itemize}

The decompressor loses synchronization if more than $W$ packets in a
row have been lost and after this event one SO packet is received.
We remark that the last condition is important: if the decompressor
never received packets, it would not be aware of the
synchronization loss.

While the decompressor is synchronized, the model tracks the number
$w$ of consecutively lost packets. Thus the model developed here is
a Markov chain in which $W+1$ states track the value of $w$, $0\leq
w\leq W$. If more than $W$ packets have been corrupted by the
channel, the decompressor may not yet be aware of the loss of
synchronization and the chain remains in a state "$W^+$" until a
packet is correctly delivered by the physical layer (note that in
this situation the channel must have transitioned from the bad to
the good state). If the packet is an IR frame, the node retrieves
synchronization and returns to the $w=0$ state. Otherwise, the
decompressor realises it has lost synchronization and moves into a
(OoS, G) state, where the G represents the channel condition. The
chain remains in the (OoS, G) until either the channel transitions
into the B state (and hence the chain moves into (OoS, B)) or an IR packet
is received, and thus the decompressor recovers the synchronization
and can return to the $w=0$ state. The decompressor has lost
synchronization when it is in either the (OoS, G) or (OoS, B) state.

\begin{figure}[!t]
\centering
\includegraphics[width=\columnwidth]{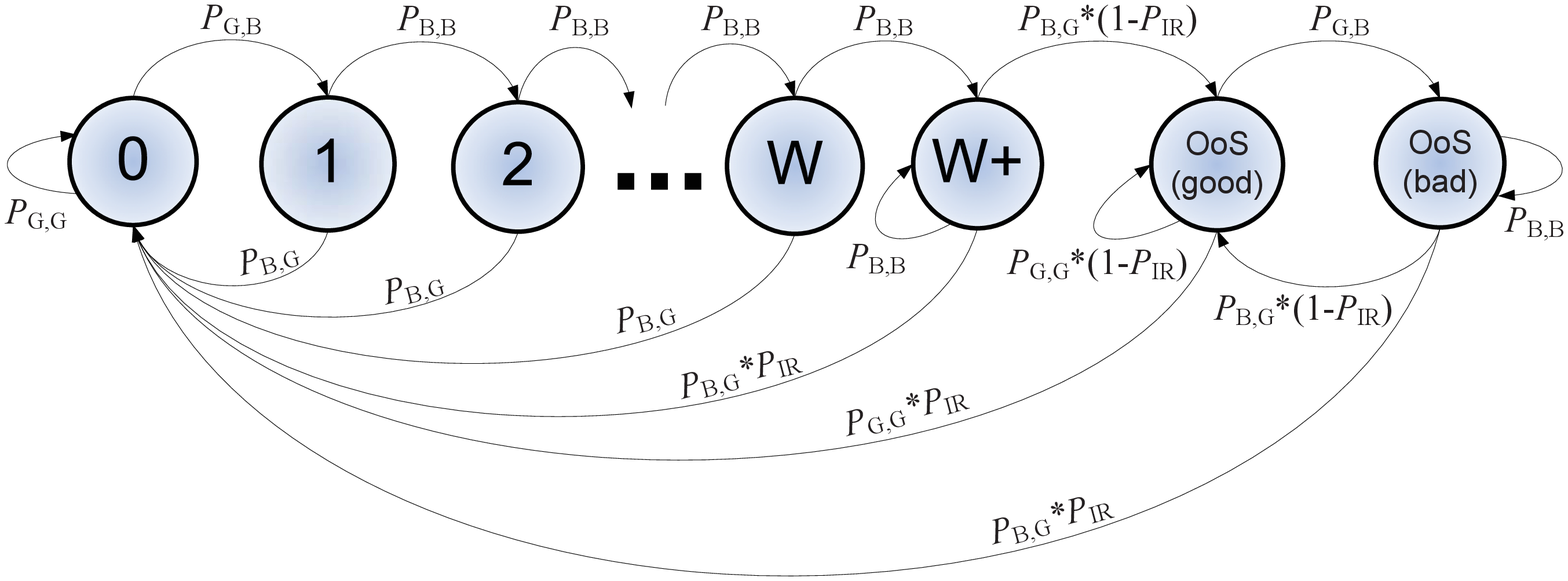}
\caption{Markov chain model for the ROHC modeling in U-mode - model 2} \label{fig:markov}
\end{figure}

Fig.~\ref{fig:markov} depicts this Markov chain model and its
transition probabilities. Let us denote by
$\pi_w,\,\pi_{W^+},\,\pi_{\text{OoS,G}},\,\pi_{\text{OoS,B}}$ the
steady state probabilities of state $w$, $W^+$ and of the two
out-of-synchronization states, respectively. The solution of the
chain is in principle straightforward, although tedious. Some steps
will be nonetheless provided. The key goal is to compute the
probability $\pi_0$ of the $w=0$ state, as all other steady state probabilities can be written as a function of
$\pi_0$. These probabilities turn out to be equal to:

\begin{eqnarray}
\pi_w       &=& P_{\text{G,B}}(P_{\text{B,B}})^{w-1}\pi_0,\,0\leq w\leq W \\
\pi_{W^+}   &=& \frac{P_{\text{G,B}}(P_{\text{B,B}})^W}{P_{\text{B,G}}}\pi_0 \\
\pi_{OoS,G} &=& \frac{P_{\text{G,B}}(P_{\text{B,B}})^W(1-P_{\text{IR}})}{P_{\text{IR}}}\pi_0 \\
\pi_{OoS,B} &=& \frac{P_{\text{G,B}}(P_{\text{B,B}})^W(1-P_{\text{IR}})}{P_{\text{IR}}}\frac{P_{\text{G,B}}}{P_{\text{B,G}}}\pi_0
\end{eqnarray}

Through these equations, the steady state solution can be computed
by means of the normalization condition (Eq.~(\ref{eq:norm_condition})):

\begin{eqnarray} \label{eq:norm_condition}
1       &=& \sum_{w=0}^W\pi_w+\pi_{W^+}+\pi_{OoS,G}+\pi_{OoS,B} \\
\pi_0   &=& \frac{1}{1+F_W+F_{\text{OoS}}} \\
F_W     &=& \frac{\sum_{w=1}^{W}\pi_w + \pi_{W^+}}{\pi_0} = \frac{P_{\text{G,B}}}{P_{\text{B,G}}} \\ 
F_{\text{OoS}} &=& \frac{\pi_{\text{OoS,G}}+\pi_{\text{OoS,B}}}{\pi_0}= \nonumber\\
        &=& \frac{P_{\text{G,B}}}{P_{\text{B,G}}}(P_{\text{B,B}})^W(P_{\text{B,G}}+P_{\text{G,B}})\frac{1-P_{\text{IR}}}{P_{\text{IR}}}
\end{eqnarray}

We shall define the probability of being out of synchronization as
$P_{\text{OoS}}$, which is equal to the probability of being in
either state (OoS,G) or (OoS,B):

\begin{eqnarray}
P_{\text{OoS}} &=& \pi_{\text{OoS,G}}+\pi_{\text{OoS,B}}= \nonumber\\
        &=& \frac{\frac{P_{\text{G,B}}}{P_{\text{B,G}}}(P_{\text{B,B}})^W(P_{\text{B,G}}+P_{\text{G,B}})\frac{1-P_{\text{IR}}}{P_{\text{IR}}}}{1 + \frac{P_{\text{G,B}}}{P_{\text{B,G}}}\left(1 + (P_{\text{B,B}})^W(P_{\text{B,G}}+P_{\text{G,B}})\frac{1-P_{\text{IR}}}{P_{\text{IR}}}\right) }\label{eq:OoS}
\end{eqnarray}

Eq.~(\ref{eq:OoS}) is not particularly insightful but can be
simplified under reasonable hypotheses in realistic settings. First
of all, it shall be assumed that $\epsilon \ll 1 \rightarrow
L_\text{B} \ll L_\text{G} \rightarrow P_{\text{G,B}} \ll
P_{\text{B,G}}$. This means that the channel does not introduce too
many errors (say, below 10\%). Hence, the denominator of
Eq.~(\ref{eq:OoS}) is very close to (just slightly larger than) 1.
Moreover, the IR timeout will be assumed to be much larger than 1
(otherwise, uncompressed packets are sent too often and the ROHC
efficiency is too low), thus $P_{\text{IR}}\simeq 0$. The numerator
can be approximated as:

\begin{eqnarray}
P_{\text{OoS}} &\simeq& \frac{P_{\text{G,B}}}{P_{\text{IR}}}(1-P_{\text{B,G}})^W = \frac{1}{L_\text{G}}\left(\frac{L_\text{B}-1}{L_\text{B}}\right)^W\text{IRT} \simeq \nonumber \\
        &\simeq& \frac{\epsilon}{L_\text{B}}\left(1-\frac{1}{L_\text{B}}\right)^W\text{IRT} \label{eq:oos_simpl}
\end{eqnarray}

The expression links the two parameters that describe the
Gilbert-Elliott channel ($\epsilon$ and $L_\text{B}$) and the two
ROHC design parameters $W$ and IRT with the OoS probability, which
is our main metric.

A natural question is how to pick the value of $W$ so that
$P_{\text{OoS}}\ll \epsilon$, that is to say, how to design the
system so that the OoS probability does not significantly worsen the
intrinsic error ratio of the channel. Let us define as $A$ the ratio
$P_{\text{OoS}}/\epsilon$ and let us set $A\gtrsim 0$ (in practice,
$A<0.1$). Hence:

\begin{equation}
   \vspace{-5pt}
W = \frac{\log\left(\frac{AL_\text{B}}{\text{IRT}}\right)}{\log\left(1-\frac{1}{L_\text{B}}\right)}
\end{equation}

\vspace{0.3cm}
If in addition $L_\text{B}\gg 1$, $\log(1-1/L_\text{B})\simeq-1/L_\text{B}$ and thus:

\begin{equation}\label{eq:B_simple}
    \vspace{-5pt}
W \gtrsim \frac{\log\left(\frac{AL_\text{B}}{\text{IRT}}\right)}{-\frac{1}{L_\text{B}}} = -L_B\log\left(\frac{AL_\text{B}}{\text{IRT}}\right) = L_B\log\left(\frac{\text{IRT}}{AL_\text{B}}\right)
\end{equation}



This equation formally proves an intuitive fact: in the
Gilbert-Elliott channel, the maximum number of packets that can be
lost in a row should be roughly proportional to the burst length
$L_\text{B}$. Similar reasoning for the IRT yields:

\begin{equation}\label{eq:IRT_simple}
   \vspace{-5pt}
\text{IRT}=AL_\text{B}\left(1+\frac{1}{L_\text{B}-1}\right)^W
\end{equation}

Eqs.~(\ref{eq:oos_simpl}), (\ref{eq:B_simple}) and
(\ref{eq:IRT_simple}) provide simple and intuitive relationships
between the system and environment parameters.

\subsection{Model 3: Multiple Flows}\label{subsec:Multiple_Flows}

The two previous models aim to represent
the ROHC protocol behavior where only one flow is considered and when only one dynamic field needs to tracked. The
goal of this last subsection is to investigate the less conventional
but relevant case of multiple RTP sources. The
focus is the derivation of a third Markov chain model
which enables to obtain more insight on the behavior of ROHC for
this specific case. The study of the effects of multiplexing together
several RTP flows on the system performance is carried out in
Section~\ref{sec:Numerical_Results}.

Let us remark that, while Model 1 and 2 apply to a variety of ROHC profiles, the model now being introduced works only for the multiplexing of several RTP/UDP/IPv4 flows together. Indeed, as explained in the next paragraphs, in this context a key problem is the compression of two dynamic fields (RTP SN and IPv4-ID). The former is encoded with the W-LSB approach, while for the latter offset encoding is assumed (i.e., the difference between RTP SN and IPv4 ID is compressed). Hence the scope of Model 3 is narrower than that of Model 1 and 2, but deals with a practically relevant problem that cannot be investigate directly with the previously developed tools. In this section, the adopted version of IP is always IPv4.

Instead of having only one packet source, the compressor multiplexes $M$ different RTP flows (where
$M>1$) which share the same channel. Let us remark that this new
model studies the evolution of a specific RTP flow, hence what is in
the end computed is the probability that ROHC goes out of
synchronization for the tagged RTP flow. The presence of multiple
flows has three main consequences.

The first consequence of having $M>1$ is that the flows are
multiplexed and therefore the RTP flow ID of the transmitted packet may
change from slot to slot. Since each flow is described by its context and the decompressor can correctly identify the context in each slot, the flow ID is always correctly recovered. The traffic model determines which flow is active in every slot.

Secondly, the robustness is not only limited by $W$ but also by
$W_o$, as stated in Section~\ref{sec:System_Model}. Thus also the latter
must be taken into account while deriving the Markov chain
for multiple RTP flows.

Third, the evolution of the channel between two consecutive IP
packets still follows a Gilbert-Elliott statistics. However, the
model tracks the behavior of a specific RTP flow among the $M$ active
ones. Hence, frames from other flows may be present between two
consecutive RTP packets of the same flow and therefore the channel
transition probabilities as observed by the tagged flow no longer
obey those of a Gilbert-Elliott model. These three elements are
studied in the next lines.

In order to devise this model, the same assumptions as in
Section~\ref{subsec:single_simplified} for the compressor and the
decompressor are adopted. The model still tracks the number $w$ of
consecutively lost packets of the same flow as well. Hence, the
chain is composed by the same states as the simplified model of
Fig.~\ref{fig:markov}. The major difference with the previous model
is the following. With a single RTP flow, the decompressor can lose
synchronization only if the difference in the SN is too large, and
hence the OoS states can be approached only from the $W^+$ state.
Fig.~\ref{fig:mult_flows} shows the example of three packets that
belong to the same flow. The first and third one (SN~=~1 and 3,
respectively) are correctly received, but the second one is deleted.
In this multi flow setting, the decompressor may lose
synchronization if the RTP SN or the IP ID go outside of their
respective interpretation window. The former case can happen only
from the $W^+$ state, but the second event implies that the number
of inserted IP packets $\Delta_{\mbox{\footnotesize{IP}}}$ between
two consecutively correctly received packets of the tagged flow
exceeds $W_o$ ($\Delta_{\mbox{\footnotesize{IP}}}$ is also depicted
in Fig.~\ref{fig:mult_flows} and in that case $w=1$ and
$\Delta_{\mbox{\footnotesize{IP}}}=4$). This may happen in fact from
any $0\leq w\leq W$ state. In order to analyze this event, an exact
modeling would also need to track
$\Delta_{\mbox{\footnotesize{IP}}}$, but this would significantly
complicate the chain and thwart the derivation of insightful
formulae, thus reducing the engineering usefulness of the model. We
prefer therefore to model the variations of $W_o$ only
statistically.

It is assumed that at each time slot one of the $M$ flows is picked
with uniform probability and that flow sends one packet. Hence the
number of slots $D_i$ between two consecutively transmitted packets
of the same flow is geometric with parameter $1/M$. At state $w$,
$\Delta_{\mbox{\footnotesize{IP}}} = w + \sum_{i=0}^wD_i$ is the sum
of $w+1$ geometric random variables follows a Pascal distribution of
parameters $w+1$ and $1/M$. The probability that
$\Delta_{\mbox{\footnotesize{IP}}}\leq W_o$ is then the cumulative
distribution function of a Pascal random variable evaluated at
$W_o-w$~\cite{Gradshteyn07}:

\begin{eqnarray}
P(w) &=& P[ \Delta_{\mbox{\footnotesize{IP}}} \leq W_o; w+1, 1/M ] = \nonumber \\
     &=& \frac{\int_0^{1/M}u^{w}(1-u)^{W_o-w}du}{\int_0^1u^{w}(1-u)^{W_o-w}du}
\end{eqnarray}
with $0\leq w\leq W$. Hence, at state $w$, if the Gilbert-Elliott
channel passes into the good state, the ROHC still retains the
synchronization and goes into state $w=0$ with probability $P(w)$,
otherwise it moves into the (OoS, G) state. We remark that if
$M=1\rightarrow\,P(w)=1,\,\forall w$ and all transition
probabilities reduce to the single flow case. Indeed, if there is
only one flow, the probability of leaving state $0\leq w\leq W$
directly into (OoS, G) is zero, as assumed in the previous section.



A final difference with respect to the single flow case is
represented by the channel transition probabilities. The channel
transitions obey the Gilbert-Elliott model for two consecutive
slots. However, multiple slots may pass between two consecutive
packets of the same flow. According to the previous discussion, the
number of slots $D$ between two consecutive RTP frames of the same
flow follows a geometric distribution with parameter $1/M$. Let us
define $\mathcal{P}$ the one step transition matrix of the
Gilbert-Elliott channel model. The average channel transition matrix
$\bar{\mathcal{P}}$, experienced by one flow, has therefore the
following form:

\begin{eqnarray}
\bar{\mathcal{P}} &=& E\left[\mathcal{P}^D\right] = \sum_{D=1}^{+\infty}\mathcal{P}^D\frac{1}{M}\left(1-\frac{1}{M}\right)^{D-1} = \nonumber \\
                  &=& \sum_{D=0}^{+\infty}\mathcal{P}^{D+1}\frac{1}{M}\left(1-\frac{1}{M}\right)^D = \nonumber \\
                  &=& \mathcal{P}\frac{1}{M}\sum_{D=0}^{+\infty}\left[\mathcal{P}\left(1-\frac{1}{M}\right)\right]^D = \nonumber \\
                  &=& \mathcal{P}\frac{1}{M}\left[I_2-\mathcal{P}\left(1-\frac{1}{M}\right)\right]^{-1}
\end{eqnarray}
where $I_2$ is the two by two identity matrix and the last equality
is guaranteed by the fact that $\mathcal{P}(1-1/M)$ has spectral
radius equal to $1-1/M<1$~\cite{Golub96}.

In conclusion, the new model is very similar to the one depicted in
Fig.~\ref{fig:markov} except that a new additional transition  between
the state $0\leq w\leq W$ and (OoS, G) must be added. The chain can be solved and the steady state and OoS probabilities
can be found. After some simplifications and under the hypothesis of small deletion rates ($\epsilon \ll1$), the OoS probability can be
expressed as seen in Eq.~(\ref{eqn_dbl_y}). We remark that $M=1\rightarrow P(w)=1,\,\forall w$, and (\ref{eqn_dbl_y}) reduces to
Eq.~(\ref{eq:oos_simpl}).

\begin{equation} \label{eqn_dbl_y}
   \vspace{-5pt}
P_{OoS} \simeq IRT\left[(1-P(0))+ \frac{\epsilon}{L_B}\left(\frac{1}{L_B}\sum_{w=1}^W \left(1-\frac{1}{L_B}\right)^{w-1}(1-P(w))-(1-P(0))+\left(1-\frac{1}{L_B}\right)^W \right)\right]
\end{equation}

%% file: Numerical_Results_Validation.tex
\section{Numerical Results}\label{sec:Numerical_Results}

\begin{figure}
\begin{center}
\includegraphics[width=0.8\columnwidth]{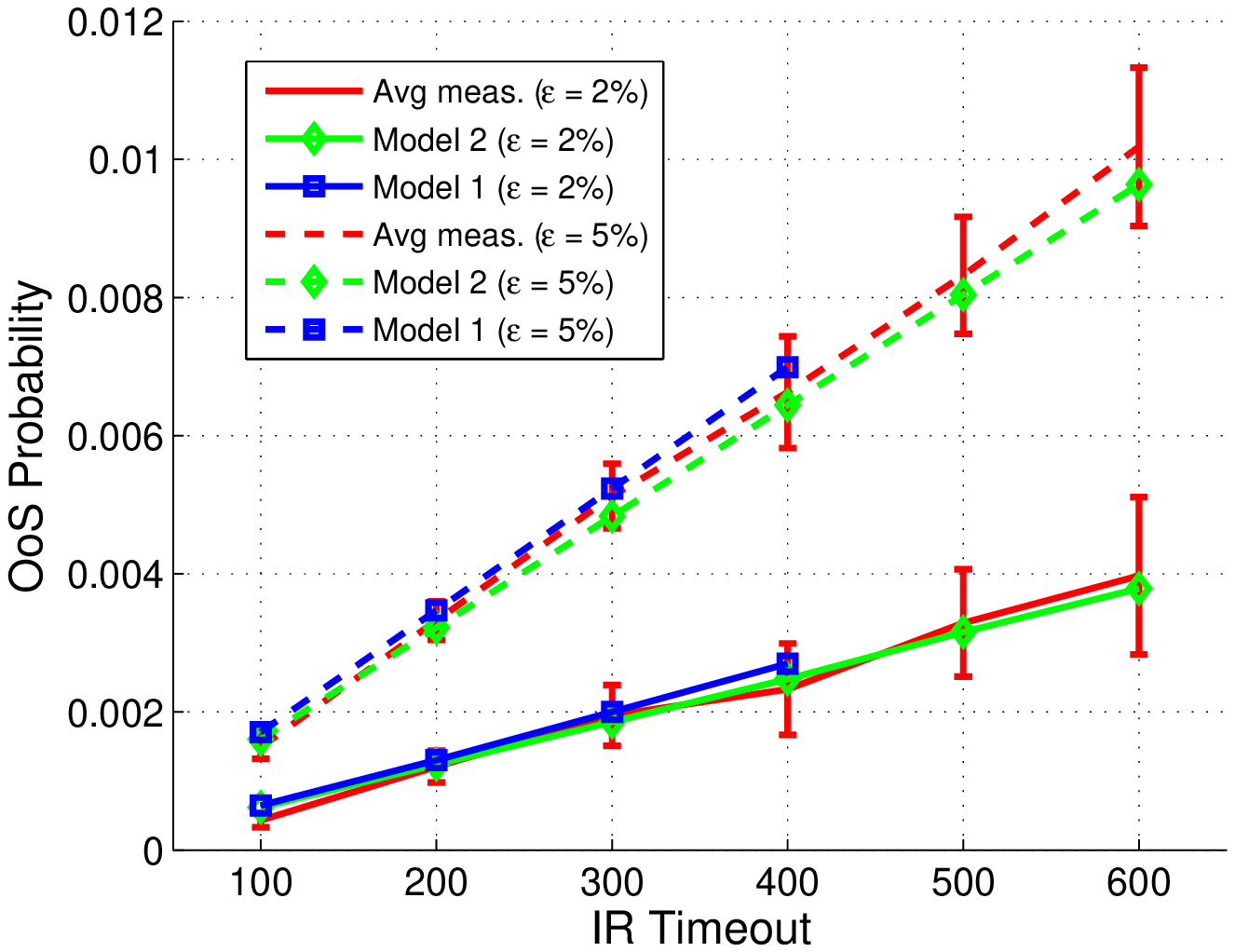}
\caption{Comparison of the OoS probability against the IRT of the theoretical models and a ROHC implementation. $W=29$, $\epsilon=2\%$, $L_B=5$.}
\label{fig:validation_graph_avg_std}
\end{center}
\end{figure}

The models have been numerically evaluated in Matlab. Unless
otherwise stated, the burst length $L_B$ and average error
probability $\epsilon$ of the Gilbert Elliott channel has been set
to 5 and 2\%, respectively, which are reasonable values for wireless
channels under moderate mobility~\cite{Badia10}. The numerical
analysis of the models starts with the single flow setting first.
The single flow models have been validated by means of a comparison
against the actual ROHC implementation from \cite{ROHC_launchpad} run in a Linux computer,
and the result is reported in Fig.~\ref{fig:validation_graph_avg_std} which shows the OoS probability against the IRT for $\epsilon=2\%,\,5\%$.
Each point reports the average of 7 simulation runs for IRT $\in\{100, 200, 300\}$, while for higher IRT values 11 simulations were carried out. Profile~1 is always employed and in each run one hundred thousand packets were transmitted, while the channel deletion pattern was generated offline by means of a two state Gilbert Elliott deletion channel model, equal to the one reported in Section~\ref{sec:System_Model}. Finally, the RTP/UDP/IP packets are generated at regular intervals. The payloads of these packets do not carry traffic, but the packet generation rate resembles that of a VoIP or video encoder. The evaluation of model 1 and 2 as well as of the real world implementation are compared. The measured performance of ROHC is depicted together with the 95\% confidence interval,while values for Model 1 can be reported up to IRT 400. Indeed, after this value, the size of the Markov chain in Fig. 4 is too large and the numerical solver in our platform does not manage to compute the solution. However, we remark that both models follow very closely the actual performance of ROHC for a rather large range of IR timeout of practical interest. Thus the models will be regarded as validated.

\begin{figure}
\begin{center}
\includegraphics[width=0.8\columnwidth]{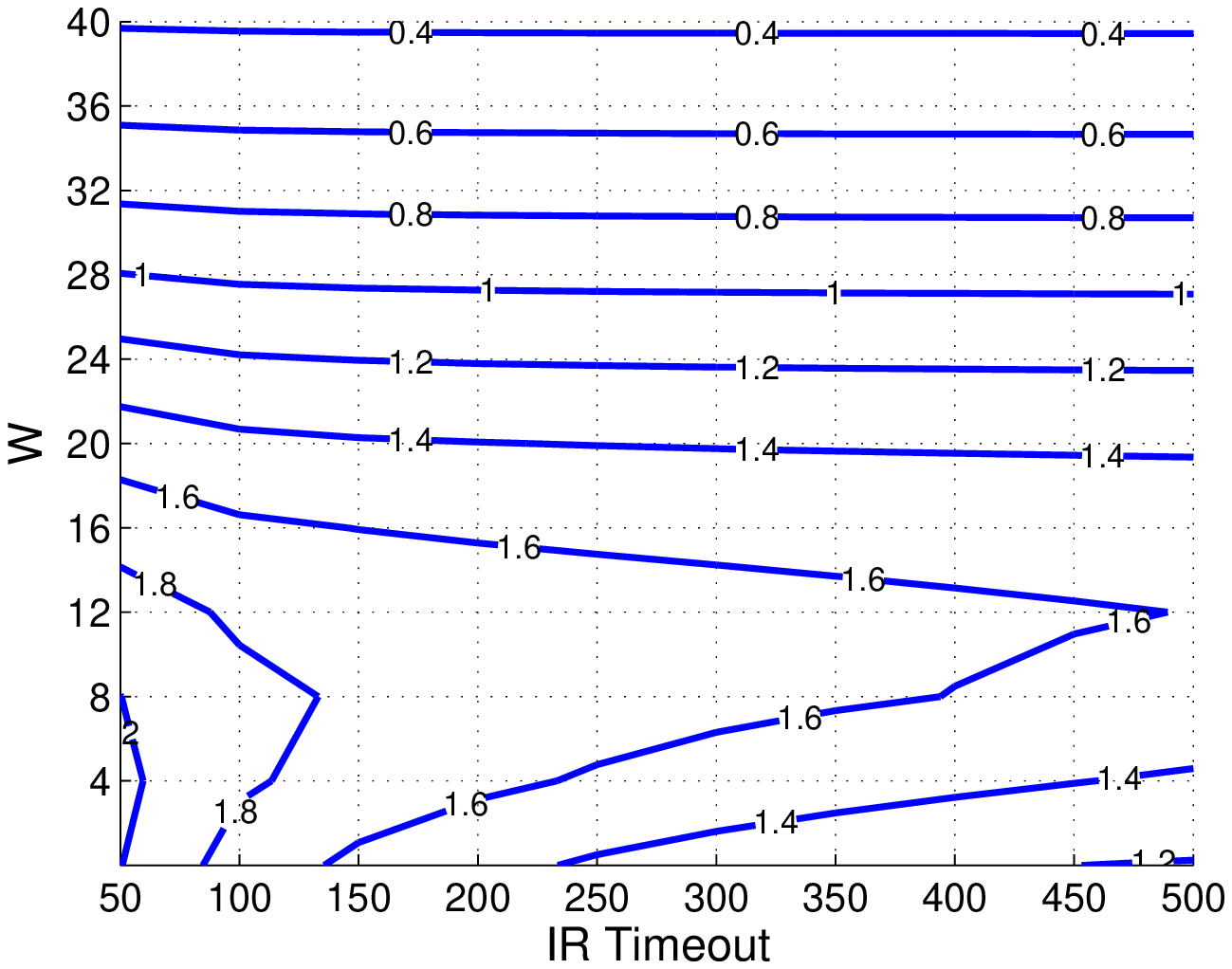}
\caption{Contour plot of the ratio between the OoS probabilites of model 2 and model 1. $\epsilon=2\%$, $L_b=5$.}
\label{fig:2_contour_likeness_2v3_4v3}
\end{center}
\end{figure}

An important point is also how well the simplified model
approximates the more sophisticated and realistic one. As stated,
the purpose of the former is not to give an accurate numerical
representation of the actual ROHC performance, but rather to foresee
the trends up to a multiplicative constant.
Fig.~\ref{fig:2_contour_likeness_2v3_4v3} shows the ratio of the OoS
probabilities as predicted by the second and first models. It can be
observed that the two models yield similar results (up to a
multiplicative factor) for a wide range of design parameters. The
multiplicative factor is mostly limited between 0.5 and 2, hence the
simplified model still yields useful first order approximations of
the OoS probability for many practically relevant values. The main
reason why the results of the simplified approach deviate from those
of Model 1 lie in the CRC check when the wraparound mechanism is
applied. In Model 1 (as in the practical ROHC implementations), when
the wraparound mechanism is employed, the decompressor may lose
synchronization even if $w<W$. This fact is ignored by Model 2 and
hence when $W$ is large the simplified approach underestimates the
OoS probability, contrary to Model 1. We remark that for $W=29$
(widely employed in practice), the two models essentially yield the
same prediction, and this is indeed the case in
Fig.~\ref{fig:validation_graph_avg_std}. Moreover, for large values of the IRT, the multiplicative factor depends very weakly on IRT and far more on $W$.

\begin{figure}
\begin{center}
\includegraphics[width=0.8\columnwidth]{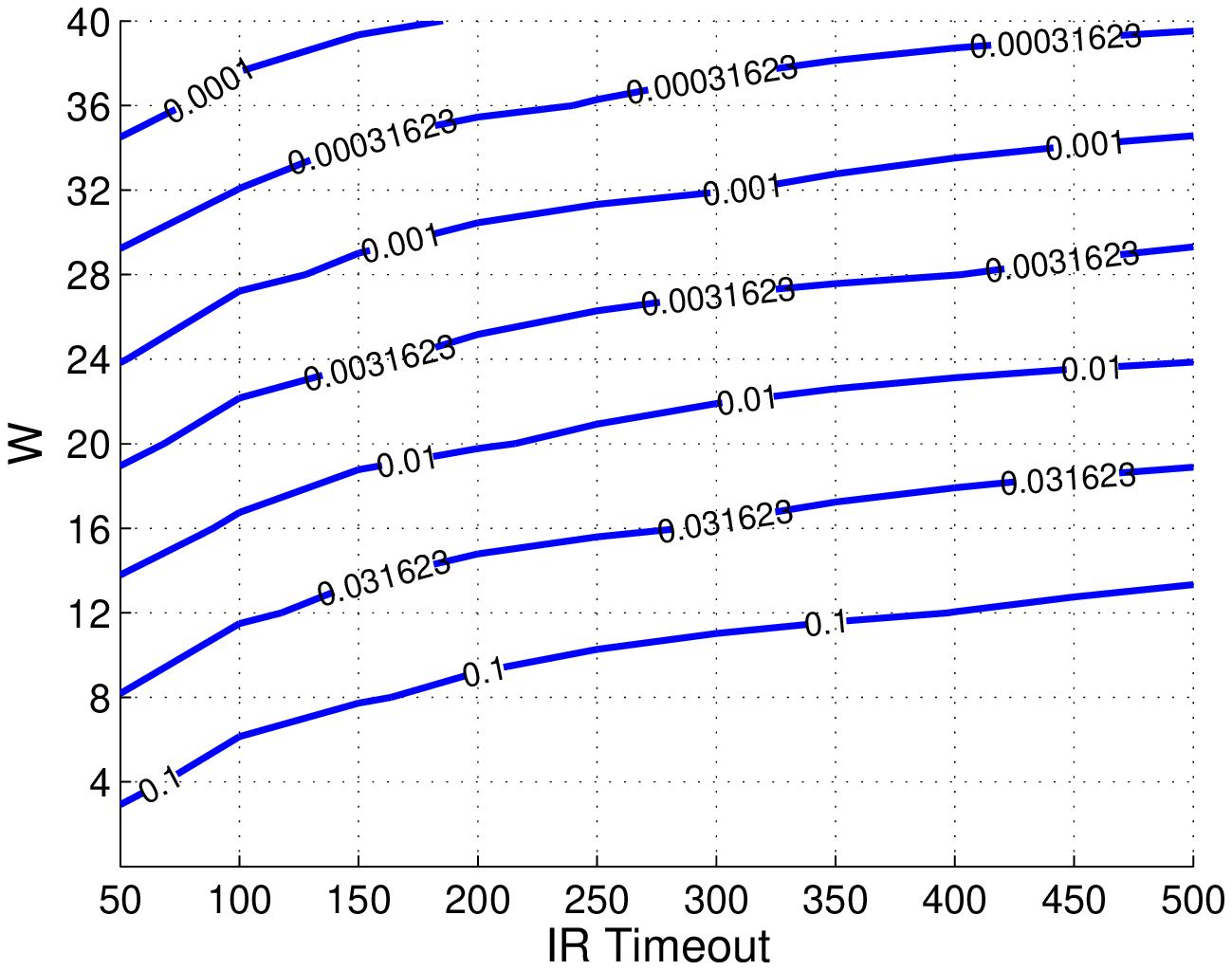}
\caption{Contour plot of the OoS probability for the second model. $\epsilon=2\%$, $L_b=5$.}
\label{fig:1_contour_model2v2}
\end{center}
\end{figure}

\begin{figure}
\begin{center}
\includegraphics[width=0.8\columnwidth]{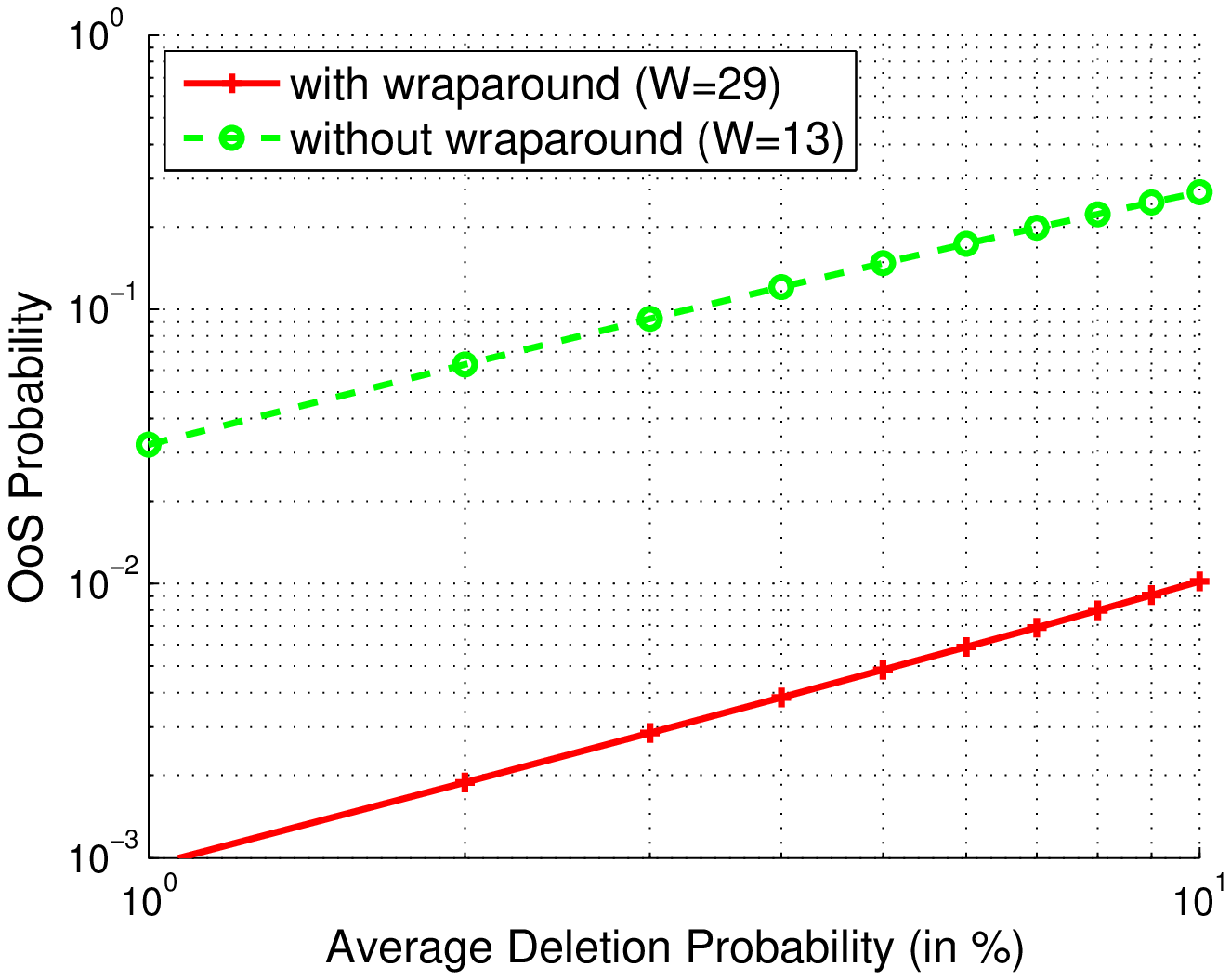}
\caption{OoS probability of model 2 against the average erasure probability $\epsilon$ with and without wraparound. $L_b=5$, IRT = 300.}
\label{fig:POoS_vs_eps}
\end{center}
\end{figure}

The previous graphs have demonstrated the validity of the proposed
models. Fig.~\ref{fig:1_contour_model2v2} represents the OoS
probability for a range of IR timeout and interpretation window $W$
for model 2. The picture shows that for $W=29$ the OoS probability
is below $0.3\%$, which is one order of magnitude smaller than the
channel error rate ($2\%$), and with this choice of $W$ ROHC does
not degrade appreciably the overall performance of the system
compared to the errors intrinsically introduced by the channel. Instead, if the wraparound mechanism was not
adopted, $W$ would drop to 13 and the OoS probability
would soar to values between 1 and 10\%; thus the ROHC mechanism
could introduce more errors than the channel does and would become the
limiting factor of the system performance. Therefore, the wraparound
mechanism is necessary to provide satisfactory performance in
correlated wireless channels. This statement is further supported by
Fig.~\ref{fig:POoS_vs_eps}, which compares the OoS probability with
and without wraparound against the average deletion probability for
$L_B=5$. The necessity of this mechanism in order to extend $W$ to
acceptable values is clear.

\begin{figure}
\begin{center}
\includegraphics[width=0.8\columnwidth]{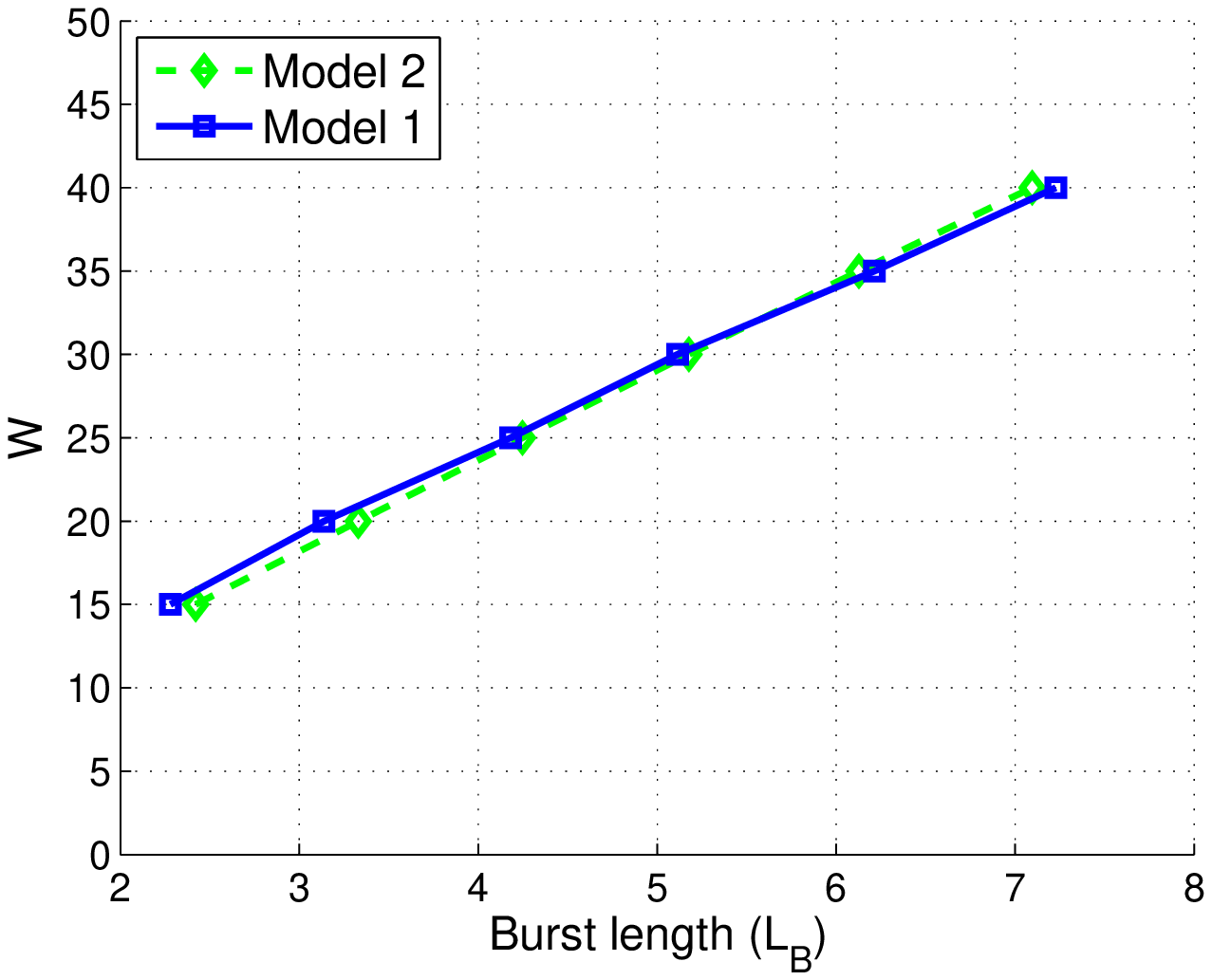}
\caption{Minimum value of $W$ to achieve an OoS probability no larger than $\epsilon/10=0.2\%$ against the burst length $L_B$. $\epsilon=2\%$, IRT = 300.}
\label{fig:5_W_vs_Lb}
\end{center}
\end{figure}

An important practical question is how the interpretation window $W$
should be tuned as the correlation time of the channel (exemplified
by $L_B$) changes, for a given target of OoS probability. In
particular, it was decided to target an OoS probability equal to
10\% of $\epsilon$, so that ROHC is not the
limiting factor in the system performance. The results are depicted
in Fig.~\ref{fig:5_W_vs_Lb}. The dependence between the error burst
length and the minimum value of $W$ is approximately linear, which
is in rough agreement with Eq.~(\ref{eq:B_simple}). The picture
suggests that the interpretation interval $W$ should be about 5.5-6
times larger than the average burst length for the given target
reliability and IR timeout, which is in agreement with the value
suggested by Eq.~(\ref{eq:B_simple}).

A similar analysis has been carried out for the IR timeout changing as a function
of $L_B$ (i.e., of the channel correlation) for a given target of
OoS probability. The results are not reported due to limits of space, but it can be shown that the choice of the IRT is
quite sensitive to the average burst length and it follows the
approximately exponentially inverse relationship of
Eq.~(\ref{eq:IRT_simple}). Moreover, the predictions of both models agree rather well
with each other, which confirms the accuracy of the simplified approach.

\begin{figure}
\begin{center}
\includegraphics[width=0.8\columnwidth]{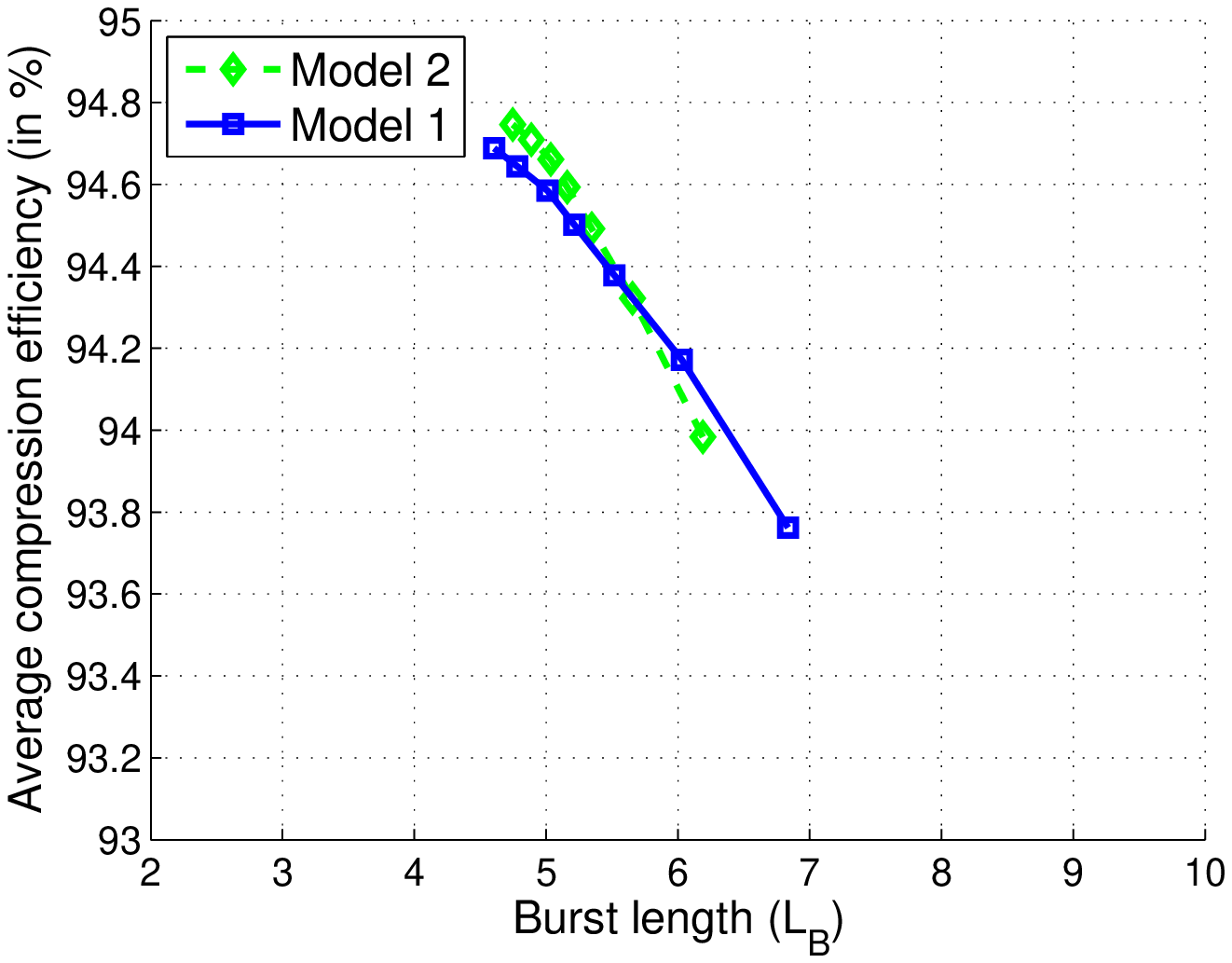}
\caption{Compression efficiency against $L_B$. $\epsilon=0.2\%$, $W=29$.}
\label{fig:6_avg_efficiency}
\end{center}
\end{figure}

While the OoS probability is a very important metric, ROHC must also
provide sufficiently high compression efficiency. Fig.~\ref{fig:6_avg_efficiency} shows the average compression efficiency when ROHC is run on IPv6 as a function of the average burst error length. This metric is defined as:

\begin{equation}\label{eq:compression_efficiency}
\mu = \frac{H_{IR}-E[H]}{H_{IR}}
\end{equation}
where $H_{IR}$ is the length of an uncompressed header and $E[H]$ is the average length of a compressed header. This ratio measures the amount of spared bandwidth due to ROHC against the bandwidth required without this compression algorithm. The target OoS probability is $0.2\%$ and the IR timeout is set to 300
as in Fig.~\ref{fig:5_W_vs_Lb}. The high compression effectiveness of ROHC
is demonstrated by the ability to shrink the header by a factor of 16-20. As
$L_B$ increases, the IRT must be decreased so as to cope with the
increased channel correlation and therefore the compression
efficiency is reduced, but a satisfactory factor of at least 15 can
be attained in all cases.

\begin{figure}
\begin{center}
\includegraphics[width=0.8\columnwidth]{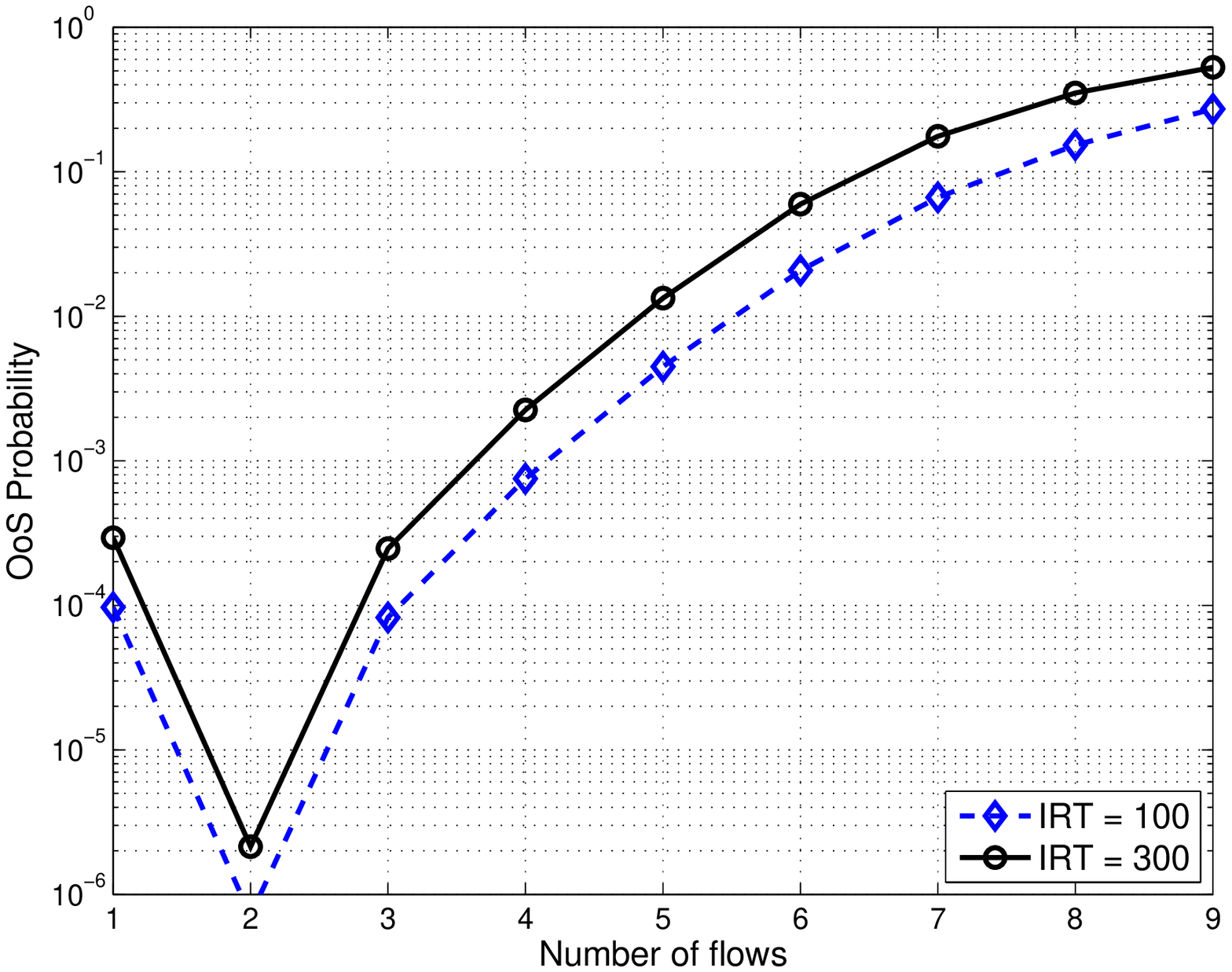}
\caption{OoS probability against $M$ for two values of IRT. $W_o=47,\,W=62$, $\epsilon=2\%$, $L_B=5$.}
\label{fig:Poos_vs_M}
\end{center}
\end{figure}

\begin{figure}
\begin{center}
\includegraphics[width=0.8\columnwidth]{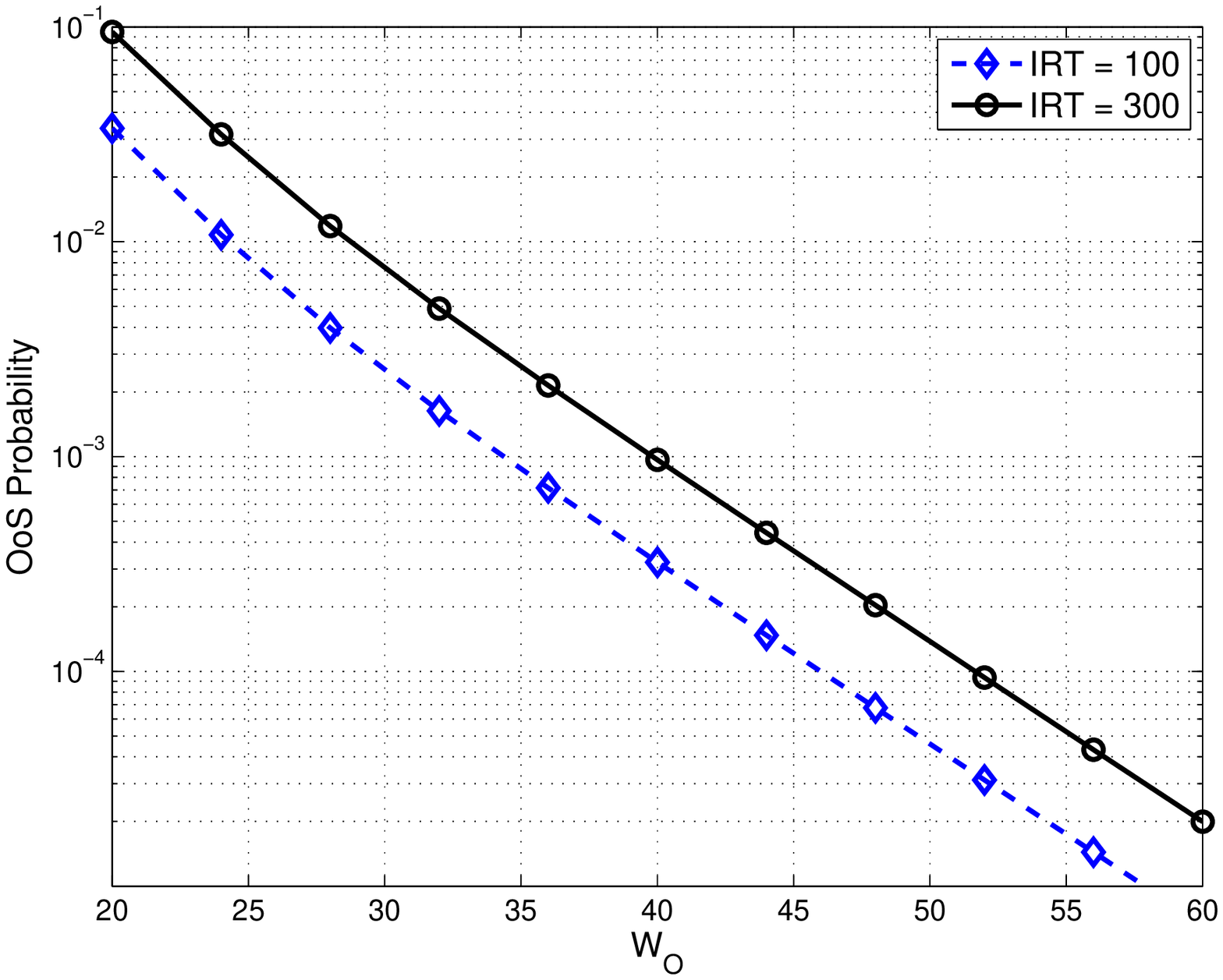}
\caption{OoS probability against $W_o$ for two values of IRT. $W=62$, $\epsilon=2\%$, $L_B=5$.}
\label{fig:Poos_vs_Wo}
\end{center}
\end{figure}

The previous results concerned the single flow case. The
last two plots explore the performance of the multi flow setting,
which is inferred through the model of
Section~\ref{subsec:Multiple_Flows}. Fig.~\ref{fig:Poos_vs_M} shows
the OoS probability against $M$ for IRT$\in\{100;300\}$. It is clear
that an excessive number of flows eventually leads to an increase of
the OoS probability, as the interpretation window of the IP
identifier will be often crossed. Indeed, for \mbox{$M\geq 3$}, the
performance degrades constantly as the number of flows is increased.
On the other hand, multiplexing three flows together slightly
improves the probability compared to having a single flow and with
$M=2$ the effect is quite dramatic. The multiplexing of more flows
together increases the time diversity and reduces the channel
correlation, hence for $M=2$ the first effect (more time diversity)
is dominant over the other consequence (increased vulnerability to
interpretation window crossings) and the mechanism is even
beneficial for small $M$. In our setting, for $M$ beyond 3 the
average error burst length becomes smaller that the average time
between two consecutive packets of the same IP flow. Hence,
additional time diversity does not help as the channel is already
sufficiently decorrelated.
Fig.~\ref{fig:Poos_vs_Wo} shows the effect of the IP identifier
interpretation interval $W_o$ for $M=3$. It is intuitive that the
OoS probability worsens with the reduction of $W_o$. The picture
shows in fact a very strong sensitivity with $W_o$ and suggests an
inversely exponential dependence of the OoS probability with $W_o$,
similar to what was observed for single IP flow profiles against
$W$.

%% file: Conclusions_Acknowledgments.tex
\section{Conclusions}\label{sec:Conclusions}

This paper has investigated a simple yet accurate model for the
Robust Header Compression Protocol in deletion channels. Our work
has shed light into the qualitative dependence of the system
behavior as a function of the channel characteristics (coherence
time and deletion probability) and the design parameters (IR timeout
and the interpretation window $W$). The model encompasses also the
practically relevant case of IP flow multiplexing and its
predictions have been widely investigated over different scenarios.
While this paper has investigated the U-mode, the introduction of a
feedback channel in the O-mode and R-mode poses interesting
questions from both a theoretical and practical point of view and
deserves investigation.

\section{Acknowledgments}\label{sec:Acknowledgments}
The research leading to these results has been partially funded by
the European Community's Seventh Framework Programme (FP7/2007-2013)
under Grant Agreement n°~233679. The SANDRA project is a Large Scale
Integrating Project for the FP7 Topic AAT.2008.4.4.2 (Integrated
approach to network centric aircraft communications for global
aircraft operations).

\appendix

The undetected error probability for the 3-bit ROHC CRC is with very good approximation 1/32, and it can be explained as follows. Since the ROHC CRC has a minimum distance of three, a possible configuration for undetected error is that the SN numbers (i.e., the systematic part) are different in three bits and the redundancy parts are equal. The most likely case to confuse the actual and reconstructed SNs is for them to be different in the three LSBs. Let us now assume that the two LSBs of the 12 context bits of the reconstructed SN are equal to 1 (which happens with probability 1/4). By definition, the true SN must be larger and the smallest number than can be added to the SN is clearly "one". The two LSBs must flip to zero, but the third bit switches as well due to the carry over. At least three bits are different and hence the CRC and the reconstructed SN may match. The CRC of this reconstructed SN is composed by 3 bits and is in general different from the original one, but there is a 1/8 chance that it is equal to the one sent in the compressed header, thus the overall CRC false negative probability of 1/32.